\documentclass[aps,pr  d,reprint, longbibliography,notitlepage,nofootinbib,nobibnotes,superscriptaddress,amsmath,amssymb,preprintnumbers]{revtex4-2}

\usepackage{graphicx} 
\usepackage{dcolumn}
\usepackage{bm}        
\usepackage{amssymb} 
\usepackage{amsmath}   
\usepackage{adjustbox}  
\usepackage{amsmath,array}
\usepackage{comment}
\usepackage{natbib}
\usepackage{MnSymbol}
\usepackage{caption}
\usepackage{subcaption}
\usepackage[colorlinks=true,linkcolor=blue,citecolor=blue, urlcolor=blue]{hyperref}
\usepackage[capitalise]{cleveref}
\usepackage[usenames,dvipsnames]{color}
\usepackage     [utf8]                  {inputenc}
\usepackage     [T1]                    {fontenc}
\usepackage     [english]               {babel}
\usepackage{epigraph}
\usepackage{etoolbox}
\usepackage{ dsfont }
\usepackage{physics}
\usepackage{float}
\usepackage{xcolor}
\usepackage{siunitx}
\makeatletter
\patchcmd{\epigraph}{\@epitext{#1}}{\itshape\@epitext{#1}}{}{}  
\newcommand*\eqsize{%
\@setfontsize\mysize{9.0}{9.0}%
    }
    \usepackage{multirow}
\makeatother

\usepackage{xfrac}
\usepackage{amsmath,amssymb}
\usepackage{esdiff} 
\usepackage{commath} 
\usepackage{bbm} 
\usepackage{braket}
\usepackage{slashed}

\newcommand{\aS}{\alpha_S}

\allowdisplaybreaks

\newcommand{\rmd}{\mathrm{d}}

\newcommand{\fm}{\mathrm{fm}}

\newcommand{\sNN}{\sqrt{s_{\mathrm{NN}}}}

\newcommand{\xT}{\mathbf{x}}

\newcommand{\sT}{\mathbf{s}}
\newcommand{\pT}{\mathbf{p}}
\newcommand{\qT}{\mathbf{q}}
\newcommand{\kT}{\mathbf{k}}

\newcommand{\Dipper}{\textsc{McDipper}}
\newcommand{\Pythia}{\textsc{Pythia}}

\definecolor{oscar}{RGB}{22, 156, 172}
\definecolor{oscarC}{RGB}{22, 156, 172}
\newcommand{\oscar}[1]{{\color{oscarC} {{\textbf{Oscar:} #1}}}}

\definecolor{LucasC}{RGB}{178, 15, 172}
\newcommand{\lucas}[1]{{\color{LucasC} {{\textbf{Lucas:} #1}}}}

\newcommand{\tauswitch}{\tau_{\mathrm{sw}}}

\makeatletter
\renewcommand\@makecaption[2]{%
  \par
  \vskip\abovecaptionskip
  \begingroup
   \small\rmfamily
    \begingroup
     \samepage
     \flushing
     \let\footnote\@footnotemark@gobble
     \@make@capt@title{#1}{#2}\par
    \endgroup
  \endgroup
  \vskip\belowcaptionskip
}
\makeatother

\begin{document}

\date{\today}

\title{3D Initial-State Dynamics across scales: A Comparative Study of saturation and string-based descriptions}
\author{Lucas Constantin}
\email{constantin@itp.uni-frankfurt.de}
\affiliation{Institute for Theoretical Physics, Goethe University, Max-von-Laue-Strasse 1, 60438 Frankfurt am Main, Germany}

\author{Oscar Garcia-Montero}
\affiliation{Instituto Galego de Física de Altas Enerxías IGFAE, Universidade de Santiago de Compostela, E-15782 Galicia, Spain}

\author{Niklas Götz}
\affiliation{Institute for Theoretical Physics, Goethe University, Max-von-Laue-Strasse 1, 60438 Frankfurt am Main, Germany}

\author{Hannah Elfner}
\affiliation{GSI Helmholtzzentrum für Schwerionenforschung, Planckstrasse 1, 64291 Darmstadt, Germany}
\affiliation{Institute for Theoretical Physics, Goethe University, Max-von-Laue-Strasse 1, 60438 Frankfurt am Main, Germany}
\affiliation{Helmholtz Research Academy Hesse for FAIR (HFHF), GSI Helmholtz Center, Campus Frankfurt, Max-von-Laue-Straße 12, 60438 Frankfurt am Main, Germany}

\begin{abstract} 
We compare the longitudinal deposition of various conserved quantities in the initial condition models of a string based (SMASH) and a saturation based ({\Dipper}) approach. SMASH has been shown to work reasonably well at lower collision energies as an initial condition for the SMASH-vHLLE hybrid approach, while \Dipper, based on the color-glass-condensate (CGC), works well in the regime of perturbative QCD. The two models are capable of providing longitudinally resolved initial conditions, which is essential for 3D hydrodynamical simulations. The goal of this study is to interface the different regions of applicability of the two models, to investigate the initial state dynamics in the intermediate energy regime. We analyze the deposition of transverse energy, charge and baryon number across a large range of collision energies (62.4 GeV to 5.02 TeV) and find that, while they are good agreement at lower energies, their energy and baryon deposition differs substantially at higher center of mass energies. 
\end{abstract}
\maketitle

\section{Introduction}
\label{sec:intro}

The primary objective of ultra-relativistic heavy-ion collisions is to investigate the properties of strongly interacting matter at extreme temperatures and energy densities. Under such conditions, Quantum Chromodynamics (QCD) predicts the formation of a deconfined state of quarks and gluons known as the quark–gluon plasma (QGP).
During the early stages of the collision, the incoming nuclei deposit not only energy and momentum into the interaction region, but also conserved charges such as net baryon number and electric charge. The subsequent evolution of these conserved quantities provides important constraints on the dynamics of the system and encodes valuable information about the structure of the initial state.

Within the standard modeling paradigm, the spacetime evolution of the hot and dense medium is described by a short equilibration stage, followed by a longer fluid-like epoch that is described by relativistic viscous hydrodynamics~\cite{Berges:2020fwq}. As the system expands and cools, it undergoes a transition back to hadronic degrees of freedom, after which hadronic transport approaches are generally employed to simulate the late, dilute stage of the collision. Because the initial state is not directly experimentally accessible, it remains one of the largest sources of theoretical uncertainty in heavy-ion phenomenology \cite{Arslandok:2023utm}. Any realistic description of the early-time medium therefore relies on assumptions about the relevant microscopic degrees of freedom and their dynamics. Constraining and understanding the initial state is thus essential for reliably extracting fundamental properties of QCD matter, such as its transport coefficients~\cite{Heinz:2013th}.

In the last decades, there has been a staggering amount of effort modeling and understanding this initial state~\cite{Mantysaari:2023gkw}. However, due to theoretical and experimental constraints, many initial condition models are formulated in two spatial dimensions and assume longitudinal boost invariance, following the Bjorken picture. Nevertheless, deviations from boost invariance are expected, even for the near-mid-rapidity regions covered in experiments at the Large Hadron Collider (LHC) and the Relativistic Heaby Ion Collider (RHIC)~\cite{Bjorken:1982qr, Pang:2015zrq}. 
The longitudinal structure of the initial state, including rapidity-dependent fluctuations and charge deposition, remains insufficiently constrained and plays an important role in phenomena such as flow decorrelations and baryon stopping \cite{ATLAS:2023rbh, BRAHMS:2003wwg}, and even to understand the initial angular momentum deposition leading to observed polarization of $\Lambda$ baryons \cite{Becattini:2007sr,Wu:2019eyi,Huang:2020dtn,Sass:2022ucj,Carrington:2025xws}. 
Additionally, even larger theoretical uncertainties affect our understanding of the intermediate collision energy regime ($\sNN \sim 10-200$ GeV) where neither purely partonic descriptions nor purely hadronic transport approaches are expected to provide a complete picture of the early dynamics. In this region, the appropriate effective degrees of freedom and mechanisms of particle production are less clearly separated than in the ultrarelativistic regime at LHC energies, or in the purely hadronic one at HADES energies. 

In this work, we compare two initial-state models based on fundamentally different microscopic assumptions and dynamical mechanisms: {\Dipper}~\cite{Garcia-Montero:2023gex}, which is rooted in gluon dynamics within the Color Glass Condensate (CGC) framework \cite{McLerran:1993ni}, and the equal-proper-time initial condition constructed from the hadronic transport code SMASH \cite{SMASH:2016zqf,Elfner:2025ojd}, which is based on hadronic interactions at low particle collision energies and string excitations at large ones. The SMASH initial condition has been used successfully within the SMASH-vHLLE hybrid model~\cite{Schafer:2021csj} to describe bulk observables at lower collision energies. While the standard CGC based  IP-Glasma + MUSIC framework is very successful at higher energies \cite{Schenke:2012wb}, the extension to 3+1D is not trivial on the initial state part. The {\Dipper} is a simplified version which yields comparable results at midrapidity for initial state observables\cite{Garcia-Montero:2023gex}. Incidentally, through the approximations within the model, it allows to connect rapidity to the initial gluon/quark distributions in a straightforward way. The main goal is to assess how these two models with completely different treatments of particle production and underlying degrees of freedom affect the longitudinal deposition of transverse energy, net-baryon number, and electric charge.
Naturally, several alternative three-dimensional initial-state models are available in the literature, such as the partonic MC-EKRT (based on collinear factorization) \cite{Niemi:2015qia, Kuha:2024kmq}, the kinetic solver AMPT \cite{Lin:2004en}, the 3D-Glauber \cite{Shen:2020jwv}, mostly based on string dynamics, and the geometry-based 3D-TRENTO \cite{Soeder:2023vdn}.


This article is organized as follows: We will introduce the two initial-state models {\Dipper} and SMASH in \cref{sec:mcdipper,sec:SMASH}, respectively. In \cref{sec:cons}, we will start our comparison with the deposition of conserved charges, particularly energy, net-baryon, and electric charges. Apart from looking at the longitudinal extension of the created fireball, we will also compare charge deposition across energies and system sizes. Subsequently, we will expand this study to geometrically sensitive quantities, where in~\cref{sec:eccentricities} we will compare initial eccentricities in the medium for both initial state models. We will summarize our findings in section \cref{sec:conclusions}. 
In \cref{sec:leadinghadrons} we will discuss the physics choices behind particle production through string fragmentation in SMASH.
\section{Initial State models}
\label{sec:ICMs}
\subsection{The {\Dipper}}
\label{sec:mcdipper}
The {\Dipper}~\cite{Garcia-Montero:2023gex,McDIPPER} is an initial state energy and charge deposition model based on the $k_T$-factorization limit of the CGC effective description of QCD \cite{Garcia-Montero:2025hys}. Using this approximation, 
energy and conserved charges are computed as moments of the single inclusive gluon and quark distributions, which are obtained via single parton production formulas computed at leading order\footnote{In this picture, gluons are produced through radiative processes, whereas at high energies, quark production enters already at leading order via the stopping of collinear projectile quarks, which are deflected from their light-cone trajectories by multiple-scatterings with the target's gluons. }(LO). 

The energy density $(e\tau)_0$ can be computed by getting the first moment of the gluon and (sea)quark distributions,
\begin{equation}
	(e\tau)_0 =\int d^2\pT~|\pT|~\left[K_g \frac{dN_{g}}{d^2\xT d^2\pT dy} + \sum_{f,\bar{f}} \frac{dN_{q_f}}{d^2\xT d^2\pT dy}\right]_{y=\eta_s}\;,
	\label{eq:EnergyWithKFactor}
\end{equation}
where in accordance with Ref.~\cite{Garcia-Montero:2023gex} we introduce the phenomenological normalization factor $K_g$ to account for normalization uncertainties and higher order corrections to  gluon production. 
Similarly, the net-charge densities $(n_f\tau)_0$ of the light flavors $f=[u,d,s]$, are the zeroth-order moments of the net-quark distribution,
\begin{equation}
	(n_f\tau)_0 =\int d^2\pT~\left[\frac{dN_{q_f}}{d^2\xT d^2\pT dy} - \frac{dN_{\bar{q}_f}}{d^2\xT d^2\pT dy}\right]_{y=\eta_s} \,.
	\label{eq:QuarkCharges}
\end{equation}
At LO, the gluon spectrum can be written as a function of transverse momentum $\pT$, momentum rapidity $y$ and transverse position $\xT$, and is given by the convolution of two gluon distributions \cite{Dumitru:2001ux,Lappi:2017skr}, 
\begin{equation}
	\begin{split}
		\frac{dN_{g}}{d^2\xT d^2\pT dy}=  &\frac{ (N_c^2-1)}{4\pi^3 g^2 N_c}  \int \frac{d^2\qT}{(2\pi)^2}   \frac{d^2\kT}{(2\pi)^2}~\frac{\qT^2\kT^2}{\pT^2}\\
		&\times~D_{1,\rm adj}(x_1,\xT,\qT)~D_{2,\rm adj}(x_2,\xT,\kT)\\
		&\times~(2\pi)^2\delta^{(2)}(\qT+\kT-\pT)\,\\
	\end{split}
	\label{eq:gluon_prod_w_uGDFs}
\end{equation}
where $D_{i,adj}(x_i, \xT,\pT)$ represents the adjoint dipole correlator of the $i$th projectile~\cite{Blaizot:2004wv,Gelis:2001da,Gelis:2001dh}, evaluated at a gluon momentum fraction $x_i$. Additionally, $N_c=3$ denotes the number of colors and $g$ is the strong coupling constant. 
What is important to note is that in this approximation of the CGC formalism, kinematics relate $x$ and rapidity in a straightforward manner, making it possible to identify the light-cone momentum fraction in the projectile and target, namely $x_{1/2}=\frac{\pT}{\sqrt{s_{NN}}} e^{\pm y}$.

Similarly, the single inclusive quark spectra is given by the deflection of collinear quarks from the projectile by the gluons in the target, represented here as the dipole operator in the fundamental representation ~\cite{Dumitru:2002qt,Dumitru:2005gt}
\begin{equation}
	\begin{split}
		\frac{dN_{q_{f}}}{d^2\xT d^2\pT dy} &= \frac{x_{1}q^{A}_{f}(x_{1},\pT^2,\xT)~D_{2,\rm fun}(x_2,\xT,\pT)}{(2\pi)^2} \\
		&+ \frac{x_{2}q^{A}_{f}(x_{2},\pT^2,\xT)~D_{1,\rm fun}(x_1,\xT,\pT)}{(2\pi)^2}\,.
	\end{split}
	\label{eq:single_quark_density}
\end{equation}
for an (anti)quark of flavor $f(\bar{f})$, where the kinematics of $x_{1,2}$ is identical to the single inclusive gluon production formula. Here, 
$q^{A}_{f}(x_{1/2},Q^2,\xT)$ are the collinear quark distributions for the two nuclei (see below). In this case, as quarks go through the initial gluon wave function, they pick up transverse momentum, and a color rotation phase. 

We model the collinear quark/antiquark distributions as an ensemble of uncorrelated partons, distributed in transverse space according to the local density of protons ($p$) and neutrons ($n$), as in 
\begin{equation}
	\begin{split}
		q^A(x,Q^2,\xT)&=q_{p}(x,Q^2)\; T_{p}(\xT) + q_{n}(x,Q^2)\; T_{n}(\xT), 
	\end{split}
	\label{eq:PDFS-pn}
\end{equation}
for $q=u,d,s$ flavor (anti)quarks, where by $T_{p/n}(x)=\sum_{i=1}^{A}~t_{p/n}(\xT-\xT_i)$ we denote the nuclear thickness of protons and neutrons in the nucleus with atomic number $A$, and where the thickness function of the individual nucleons is set to 
\begin{equation}
	t(\xT)=\frac{1}{2\pi B_G}\exp\left[-\frac{\xT^2}{2\,B_G}\right]\,,
	\label{eq:thickness}
\end{equation}
where the nucleon size, $B_G=0.156 \fm^{2}$, is determined through fits to the HERA data \cite{Kowalski:2003hm,Rezaeian:2012ji}. The thickness function of an individual nucleon is then normalized as $\int d^2\xT~t(\xT)=1$ such that $\int d^2\xT \,[T_{p}(\xT) +T_{n}(\xT)]=A$ for each nucleus. 
While sub-nuclear fluctuations have been recently implemented in the {\Dipper}\,\cite{Garcia-Montero:2025bpn}, in this work we will only use geometric fluctuations at the level of nucleons for a better comparison with SMASH. Additionally, we use isospin symmetry to obtain the neutron PDFs, as  these are not well constrained. The neutron PDFs are then set to the proton PDFs after using  $u\leftrightarrow d$. 

\subsubsection{Saturation model: IP-Sat}
The dipoles in \cref{eq:gluon_prod_w_uGDFs,eq:single_quark_density} depend on the color-field configurations assumed for the incoming targets. The correlations of these configurations are generally quite complex, but can be phenomenologically modeled. Such is the case for the impact-parameter dependent saturation (IP-Sat) model~\cite{Kowalski:2006hc,Kowalski:2003hm}, which will be the saturation dipole model used in this work to compute the initial states in the {\Dipper}, where the dipole is given by
\begin{equation}
	D_{\rm fun}(x,\xT,\sT)=\exp\left[-\frac{\pi^2\sT^2}{2\,Nc}\aS(\mu^2)\, xg(x,\mu^2)\, T(\xT)\right]\,.
	\label{eq:IPSatDip}
\end{equation}
The function $g(x,\mu^2)$ is the gluon parton distribution (PDF) function initialized at a scale $\mu_0$ via the parametrization 
\begin{equation}
	xg(x,\mu^2_0)= a_g\,x^{-\lambda_g}\, (1-x)^{5.6}\,.
\end{equation}
and evolved to further scales, $\mu$, via the Dokshitzer-Gribov-Lipatov-Altarelli-Parisi (DGLAP) equation. Additionally, in the IP-Sat model the scale at which the coupling and the gluon parton distribution function (PDF) is evaluated using $\mu^2 = \mu^2_0 + C/\sT^2$. The parameters for this evolution and saturation model are obtained from the fit given in \cite{Rezaeian:2012ji}. 

\subsection{SMASH Initial Conditions}
\label{sec:SMASH}
\begin{figure}[t]
	\centering
	\includegraphics[width=0.45\textwidth]{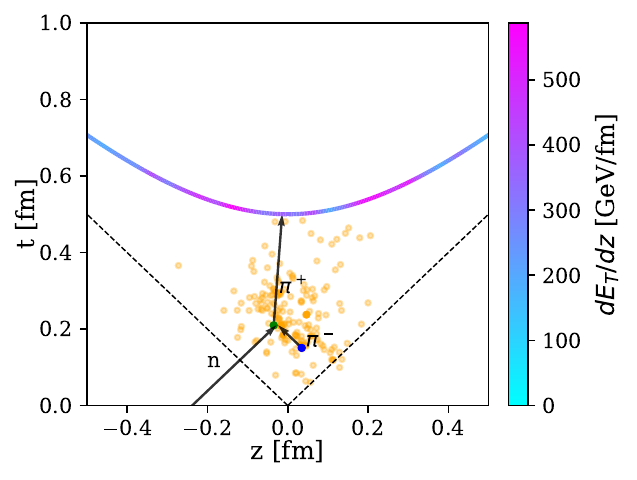}
	\caption[t-z-diagram]{$t-z$ diagram of the SMASH initial conditions. The figure shows an example Au-Au collision at $\sqrt{s_{\mathrm{NN}}}$=62.4 GeV. The orange dots represent the interaction points between particles in the evolution of SMASH. For demonstration, the collision of an initial neutron with a $\pi^-$ is shown. The string excitation is handled in {\Pythia} and one of the resulting hadrons ($\pi^+$) is highlighted. Because of the reduced cross section during the formation time, the newly formed hadrons free stream until they pass the hypersurface of $\tauswitch$ = 0.5fm, where they get removed.} 
	\label{fig:isotau}
\end{figure}
The initial conditions for the SMASH-vHLLE hybrid model \cite{Schafer:2021csj} are constructed from the hadronic evolution of SMASH, which propagates hadrons based on the relativistic Boltzmann equation. The SMASH transport evolution starts from the initial state for the colliding nuclei based on Woods-Saxon distributions or external configurations. When the nucleons collide, {\Pythia}~\cite{Bierlich:2022pfr} is employed to model the inelastic interactions by fragmenting the resulting color strings into hadrons\footnote{By default, SMASH uses the monash tune above collision energies of 200 GeV. However, in this study, the default Pythia tune is used at every collision energy to avoid discontinuities in the energy scans.}. These hadrons are propagated in SMASH with a formation time during which their interactions are suppressed, and only afterwards do they re-interact with their full hadronic cross-sections. To construct the initial state for the hydrodynamical evolution, a hypersurface of constant proper time is defined using the passing time
\begin{equation}
	\label{eq:tau_0}
	\tauswitch = \frac{R_p + R_t}{\sqrt{\left(\frac{\sqrt{s_{\mathrm{NN}}}}{2m_{\mathrm{N}}}\right)^2 -1}}
\end{equation}
of the two nuclei. In practice, a lower bound of $\tau_{\mathrm{sw}}$~=~0.5~fm is applied to ensure a sufficiently long non-equilibrium evolution. Every particle that crosses this hypersurface is removed from the hadronic evolution. 
To start the hydrodynamical evolution, the conserved quantities are distributed over the grid cells from the list of particles by applying a Gaussian smearing kernel that is Lorentz-contracted in the longitudinal direction
\begin{equation}
	\Delta P^{\alpha}_{ijk}\;=\;P^{\alpha}\,C\,\exp\!\left[-\,\tfrac{\Delta x_i^2 + \Delta y_j^2}{R_\perp^2}\,-\,\tfrac{\Delta\eta_k^2}{R_\eta^2}\,\gamma_\eta^2\,\tau_0^2\right],
	\label{eq:gaussian_smearing}
\end{equation}
where $\gamma_\eta = \mathrm{cosh}\left( y_p - \eta_s\right)$ is the Lorentz factor of a particle with momentum rapidity $y_p$ as seen by an observer moving with velocity corresponding to $\eta_s$. The smearing parameters $R_{\perp}$ = 0.63~fm and $R_{\eta}$ = 1.69~fm were inferred from experimental data via the Bayesian analysis performed in~\cite{Gotz:2025wnv}.

%

For this study, we use the Wood-Saxon parameters used in the {\Dipper}, which are sourced from \cite{dEnterria:2020dwq}. The nuclear radius and diffusiveness for Au are $R_{\mathrm{Au}} = 6.38$ fm and $a_{\mathrm{Au}} = 0.535$ fm, respectively. For the lead nucleus, these are given by $R_{\mathrm{Pb}} = 6.67$ fm and $a_{\mathrm{Pb}} = 0.54$ fm.  
\section{Longitudinal Deposition of Conserved Quantities}
\label{sec:cons}
\begin{figure}[t]
	\centering
	\includegraphics[width=\linewidth]{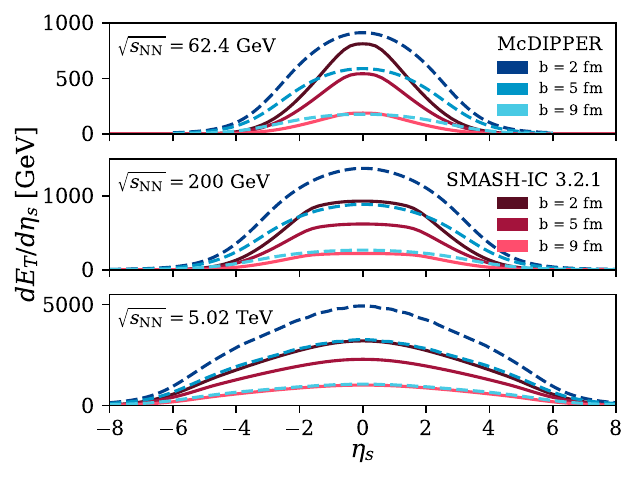}
	\caption[dE/d$\eta_s$]{Longitudinal deposition of transverse energy density $dE_{T}/d\eta_s$ for different collision energies from {\Dipper} and the SMASH-IC. Shown are multiple impact parameter classes. Both models show the results at $\tau_{sw}$ = 0.5~fm. The SMASH results were obtained by smearing the parameters in accordance with section~\ref{sec:SMASH}. The panels show different collision energies: $\sNN=$62.4, 200, 5020~GeV.} 
	\label{fig:dE_deta}
\end{figure}
In this section, we present a comparison of the initial condition models mentioned above: the saturation-based model {\Dipper} and the hadronic transport approach SMASH\footnote{SMASH 3.2.1 was used for hadronic transport based ICs, while the saturation ICs were produced using {\Dipper} 1.3~\cite{McDIPPER}. Unless stated otherwise, all model parameters are kept at their current default values.}. Results are shown for nucleus–nucleus collisions at center-of-mass energies of $\sqrt{s_{NN}}=62.4, 200 $ and $5020$ GeV. We want to compare events with different geometries, for which we have chosen three settings, for fixed impact parameters $b=2,5, 9$ fm. These choices correspond qualitatively to central, mid-central, and peripheral collisions. 
\begin{figure*}[t]
	\includegraphics[width=0.85\textwidth]{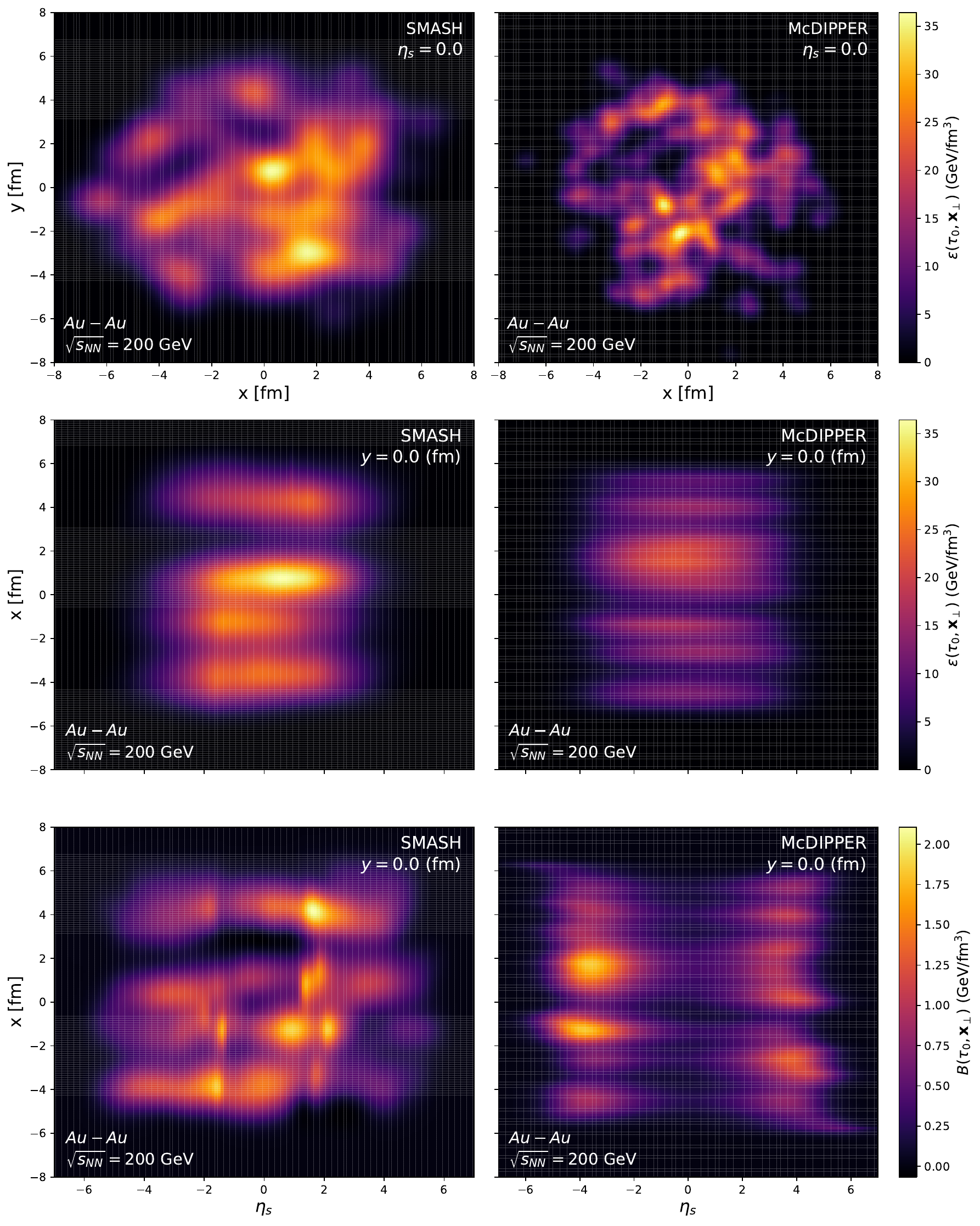}
	\caption{Slices in the transverse plane (top) and the x-$\eta_s$ plane (middle and bottom) of the transverse energy density and the baryon density. The setting is a Au-Au collision at $\sNN=200$~GeV and b=2~fm at a time of $\tau_{sw}$ = 0.5~fm.}
	\label{fig:density_slices}
\end{figure*}
All comparisons are performed using event-by-event fluctuations and presented as ensemble averages over the specified impact-parameter classes.
We begin our comparison with the deposited transverse energy $dE_\perp/d\eta_s$. While entropy density is the quantity thought of as the proxy for particle production, these two models possess very different degrees of freedom, energy (charge) deposition mechanisms and frameworks. To compare them, we would need to use different definitions for the entropy density, making the comparison non-transparent. For these reasons, we will compare the deposited transverse energy and deposited baryon and electric charges, which are directly comparable within the models.

\subsection{Creating a medium: the deposition of energy}
\begin{figure}[t]
	\centering
	\includegraphics[width=0.5\textwidth]{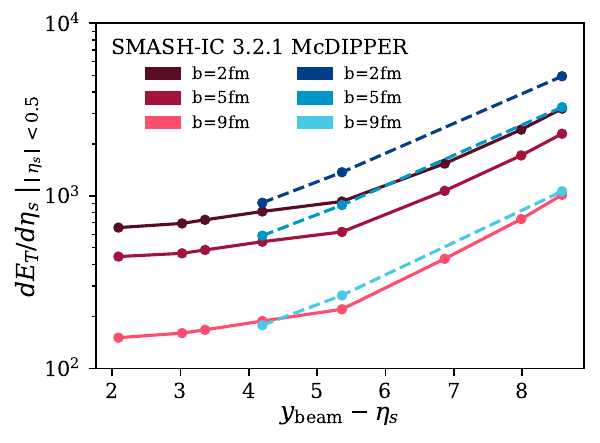}
	\caption{Transverse energy deposition at midrapidity as a function of rapidity shift $y_{\mathrm{beam}} - \eta_s$ in the SMASH-IC and {\Dipper} at $\tau_{\mathrm{sw}}$ = 0.5~fm. The data points in SMASH correspond to the collision energies 7.7, 19.3, 27, 62.4, 200, 900, 2760 and 5020~GeV.}
	\label{fig:rapidity_shift_energy}
\end{figure}

The rapidity dependence for the transverse energy density $dE_T/d\eta_s$ is presented in \cref{fig:dE_deta}. It is interesting to note that, while the two models seemingly exhibit notably different behavior for transverse energy deposition, some similarities also emerge. At lower collision energies ($\sNN=62.4$~GeV), SMASH and {\Dipper} produce similar energy densities at midrapidity. However, the rapidity structure of the deposited energy differs: the {\Dipper} 's energy profile extends considerably further toward forward and backward rapidities, creating a much broader longitudinal profile, while SMASH produces a sharper depletion of particles at forward and backward rapidities.
As the collision energy increases, {\Dipper} deposits substantially more transverse energy than SMASH. However, the profiles become more similar in rapidity shape, differing mostly by their total normalization.

It is interesting to understand $dE_T/d\eta_s$ at midrapidity as an excitation function, i.e., as a function of the beam rapidity shift $y-y_{beam}$. This variable is chosen because excitation functions regarding the deposited net-proton and charge distributions have been reported by experimental groups, and one can think of it as a proxy for energy density, where 
\begin{equation}
	y_{beam} \approx \log\left(\frac{\sNN}{m_\mathrm{N}}\right).
\end{equation}
In \cref{fig:rapidity_shift_energy} we show the dependence of the midrapidity $dE_T/d\eta_s$ on the rapidity difference $y_{beam} - \eta_s$. Results from our two IC models, SMASH-IC and {\Dipper}, are compared across the three aforementioned IP classes $b =2,5,9$ fm. The {\Dipper} generally predicts higher $dE_T/d\eta_s$ deposition than SMASH-IC, with this difference becoming more pronounced at larger rapidity separations. However, it is interesting to note that at large rapidity separations, the shape of the $dE_T/d\eta_s$ profiles and their $\sNN$ dependence become similar, despite occurring at significantly different absolute values.

Finally, we can qualitatively compare the behavior of energy deposition in the transverse plane. For this, we show in \cref{fig:density_slices} slices of the energy and baryon density of a single Au-Au collision at $\sNN=200$~GeV. The upper two plots show the transverse plane at midrapidity, showing that the two models exhibit different spatial structures, where the {\Dipper} naturally exhibits a more granular structure with multiple isolated hot spots scattered across the collision region, while the particlization process in SMASH produces smooth energy density distributions. It is important to note that while it is true that the SMASH results depend on the smearing parameters\footnote{The {\Dipper} indeed also depends on such a parameter via the definition of the nucleon thickness-function. However, that parameter, $B_G$ is previously fit to HERA data \cite{Kowalski:2003hm,Rezaeian:2012ji}, and hence it is not treated as a \textit{true} free parameter.}, if one varies $R_\perp$ around its nominal value, while remaining consistent with the values extracted from other Bayesian analyses in the community \cite{Giacalone:2023cet,Nijs:2020roc,Nijs:2020ors}, the relative difference in the behavior of these transverse structures between SMASH and the {\Dipper} will not change qualitatively. Nonetheless, $R_\eta$ is not very well constrained by data \cite{Soeder:2023vdn} and more variation will still be allowed. On the other hand, the longitudinal structures in the {\Dipper} stem directly from the $x$ dependence of the gluon distributions, which in turn result from the QCD evolution equations~\cite{Garcia-Montero:2025hys}. While there is freedom on choosing the initial model (initial dipole at some reference $x=x_0$), the choice must always be taken so that the model can already describe the inclusive dipole cross-section for the $\gamma^*p \to X$ in DIS experiments, i.e. HERA data. 
 
 \subsection{Baryon stopping}
The deposited conserved charge densities, baryon and electric charge, differ significantly between the two models. SMASH retains more baryons at midrapidity than the {\Dipper}. For the latter, the maximal deposition of the charges, the "peaks" of the distributions, appears at more forward/backward rapidities. Qualitatively speaking, there is an effective shift of around two units of spatial rapidity between the models, with this gap becoming smaller in the high energy regime. Additionally, for the whole range of energies probed, the shape of these peaks is also very different, with the forward (backward) SMASH peak exhibiting a larger skewness towards the  forward (backward) rapidities than the {\Dipper}. At very high collision energies ($\sqrt{s_{\text{NN}}} = 5.02$~TeV), SMASH exhibits a plateau of nearly constant baryon density rather than the expected quasi-vanishing deposited charge, a consequence of how leading hadrons are treated via \Pythia\ string excitations (discussed in detail in Appendix~\ref{sec:leadinghadrons}). 

Integration over rapidity reveals that generally, SMASH deposits a higher total net-charge and net-baryon number than {\Dipper}  for the same energy and IP-class events. This difference arises due to the fact that SMASH deposits the charge and baryon number from all participant nucleons, whereas in \Dipper, quark deposition probabilities depend on both the quark flavor probability in the projectile and the scattering probability, yielding lower net baryon and charge numbers than expected from the number of participants. 

\begin{figure*}[t]
	\centering
	\begin{minipage}{0.49\textwidth}
		\centering
		\includegraphics[width=\textwidth]{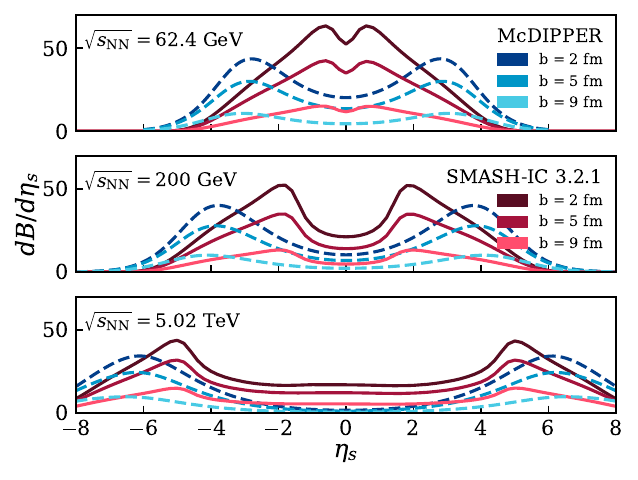}
	\end{minipage}
	\hfill
	\begin{minipage}{0.49\textwidth}
		\centering
		\includegraphics[width=\textwidth]{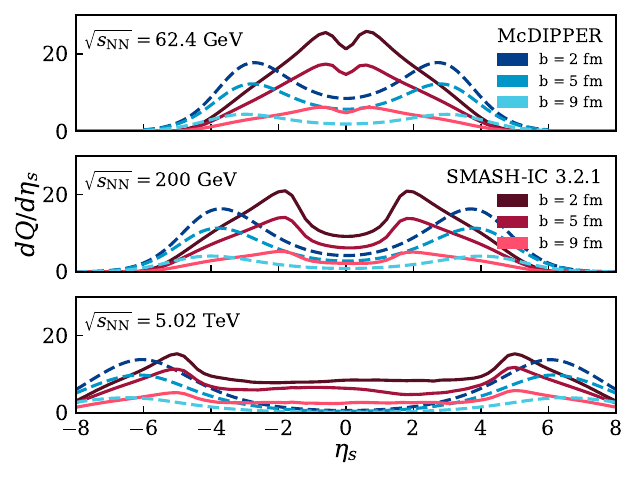}
	\end{minipage}
	\caption{Longitudinal baryon (a) and charge (b) deposition from {\Dipper} and SMASH-IC at $\tau_{\mathrm{sw}}$ = 0.5~fm as a function of spacetime rapidity $\eta_s$ for different impact parameters $b$. The panels show different collision energies: $\sNN=$62.4, 200, 5020~GeV.}
	\label{fig:QBdeposition}
\end{figure*}

\begin{figure*}[t]
	\begin{minipage}{0.49\textwidth}
		\includegraphics[width=1\textwidth]{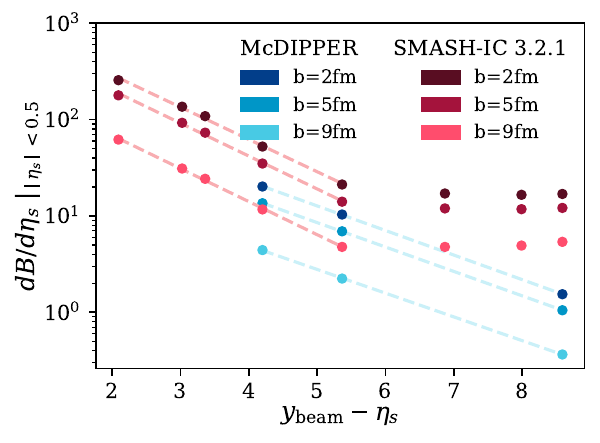}
	\end{minipage}
	\hfill
	\begin{minipage}{0.49\textwidth}
        \includegraphics[width=1\textwidth]{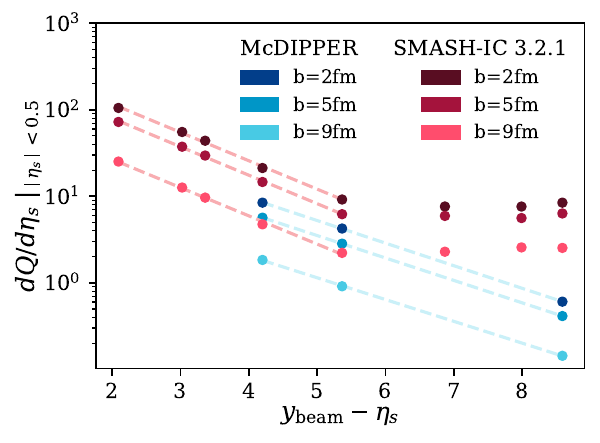}
	\end{minipage}
    \caption{Baryon deposition at midrapidity as a function of rapidity shift $y_{beam} - \eta_s$ in the SMASH-IC and {\Dipper} at $\tau_{sw}$ = 0.5~fm. The data points from SMASH correspond to the collision energies 7.7, 19.3, 27, 62.4, 200, 900, 2760 and 5020~GeV. The dashed lines represent linear fits to the results at 62.4 to 5020 GeV in the \Dipper and 7.7 to 200 GeV in SMASH. For the latter, the higher energies were excluded, because the model does not show the expected linear decrease.} 
    \label{fig:rapidity_shift}
\end{figure*}

\begin{figure}[t]
	\centering 
	\includegraphics[width=0.45\textwidth]{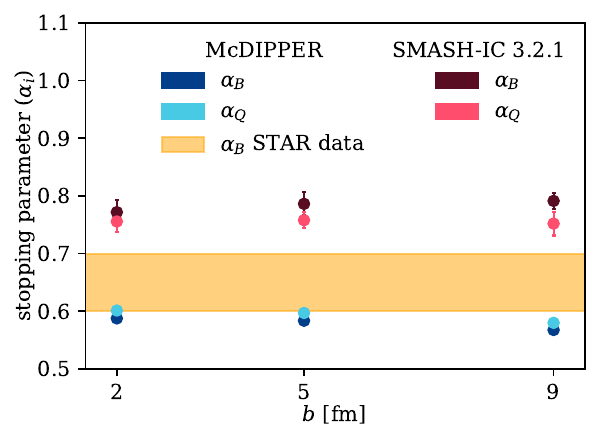}
	\caption[Baryon and Charge stopping parameter]{Baryon and Charge stopping parameters $\alpha_i$ as a function of impact parameter from SMASH-IC and {\Dipper}. Results are obtained from the linear fits shown in Fig~\ref{fig:rapidity_shift}. The yellow band shows the experimental data from the STAR collaboration~\cite{STAR:2024lvy}. Since there is no clear correspondence between centrality and impact parameter, we present it as a band and it is meant to serve as a guideline. Also, it is noteworthy that the energy range of the experimental data (7.7-200~GeV) does not necessarily correspond to the one used in the linear fits for the models.}
	\label{fig:stopping_parameters}
\end{figure}

An interesting observable is to plot the midrapidity deposition slice as an excitation function. It is a well-known behavior, both in theoretical models and in experimental data \cite{STAR:2024lvy}, that the deposited baryon number decreases exponentially with increasing rapidity shifts (increasing $\sNN$)
\begin{equation}
	\left.\frac{\rmd Q_i}{\rmd\eta_s}\right|_{|\eta_s|<0.5} \propto e^{\alpha_i \,(\eta_s - y_{\rm beam})}.
	\label{eq:exp_ybeam}
\end{equation}
In \cref{fig:rapidity_shift} we show the baryon and charge deposition at midrapidity as a function of the rapidity shift $\eta_s - y_{\rm beam}$. The dashed lines represent the linear fit of the data. One can see that SMASH and {\Dipper} follow the expected exponential in their expected validity domains, SMASH for low and intermediate energies, and the {\Dipper} for the ultra-relativistic region. In the latter region, the SMASH-IC experiences a breakdown of the exponential behavior and one can observe that the baryon density plateaus, remaining approximately constant (slightly increasing) with collision energy. This is due to the way that SMASH deals with hadronic production through string excitations in {\Pythia}. 
In short, SMASH differentiates between soft and hard string excitations. Hard string excitations are employed in high energy collisions where perturbative QCD is applicable and are handled entirely by {\Pythia}'s multiparton interaction (MPI) framework. Soft string excitations, on the other hand, are employed in the low to intermediate energy regime. Here, the essential difference is that, based on the momenta and flavors of the colliding hadrons, SMASH only uses {\Pythia} to fragment the string into hadrons. As the collision energy rises, the cross section for the hard string excitations increases.  
Another change, specifically at 200~GeV, is a change in the formation time of newly produced hadrons in SMASH. Whenever {\Pythia} is employed to handle a hadronic interaction, the cross sections of newly produced particles are temporarily suppressed in the evolution of SMASH during their formation time. Below $\sNN$=200 GeV, this suppression is gradually lifted, with the cross section increasing continuously over time. Above 200 GeV, however, the behavior changes to a step function, where the cross section remains fully suppressed until the formation time has passed and then instantaneously reaches its full value. It is these transitions at high energies that eventually leads to the change in behavior that can be seen in SMASH. 
Furthermore, at higher energies, the treatment of leading hadrons of the strings plays an important role in the stopping of baryons, as we explore in Appendix~\ref{sec:leadinghadrons}. For more information on the process types and how their cross sections are realized inside SMASH, we refer to \cite{Mohs:2019iee}, where this string based approach was successfully applied to model baryon stopping for a vast range of lower collision energies. 
Using \cref{eq:exp_ybeam} to fit the simulated data, one can extract the baryon and electric charge stopping parameters, $\alpha_B$ and $\alpha_Q$, respectively.
The resulting parameters for the two models are presented in \cref{fig:stopping_parameters} for different IP-classes. The yellow band shows the baryon stopping coefficient as measured by the ATLAS collaboration~\cite{STAR:2024lvy}. This band is meant as a guide, since the comparison needs to be performed by selecting for centrality, and not in fixed IP averages. While the {\Dipper} seems to agree with the experimental data in this plot, the reader is reminded that previous work \cite{Garcia-Montero:2024jev} locates $\alpha_B$ in the upper edge of this band. This is not a contradiction, and can be understood using the simplified formula extracted in \cite{Garcia-Montero:2024jev}, where we can focus on the forward charge deposited from the valence quarks $q_{v,f}$ of flavour $f$ of nucleus A scattering off the saturated target B 
\begin{equation}
	  \frac{\rmd n_{f,A\to B}}{ \rmd^2 \xT \rmd \eta_s}= \bigg\langle x_1 \, q_{v_f,A}\!\left(x_1,\, Q_{s,B}^2\!\left(x_2, T_B(\xT)\right)\right)T_{A}(\xT)\bigg\rangle_e,
\end{equation}
where $\langle\cdot\rangle_e$ corresponds to event averaging. Here, $Q_s,A$ corresponds to the saturation scale of the target, and characterizes the typical transverse momentum of its gluons. Normally, as in the case of the most used saturation models (for example the IP-Sat, McLerran-Venugopalan and the Golec-Biernat-Wustoff (GBW)) , $ Q_s,A \sim T_A$ 
\begin{equation}
 		\frac{\rmd n_{f,A\to B}}{ \rmd^2 \xT \rmd \eta_s}\sim \int  \big\langle T_A(\xT)T^{\gamma}_B(\xT) \big\rangle_e
\end{equation}
where $\gamma$ is a non-trivial exponents depending on the power-law-like large-$x$ behaviour of the valence quark PDF, $q_{v,f}$, as well as on the effective energy dependence of the gluon distributions (dipole) due to the QCD evolution. It is easy to see that due to the non linearity given in the saturation picture, IP and centrality classes will give relatively different results. 
On the other hand, SMASH overestimates the stopping parameter in the lower energy region, but relatively less than pure {\Pythia}, see Ref.~\cite{Lewis:2022arg}. A full, centrality selected, simulation is needed for a proper comparison. However, to compare them to each other and to data, embedding the models into a full hybrid model is needed. While this has already been performed, such a comparison is beyond the scope and main point of this paper. 

\section{Event eccentricities}
\label{sec:eccentricities}
One of the key observables for characterizing the properties of the QGP is the anisotropic flow of final-state particles, which arises from the collective hydrodynamic response of the medium to the spatial anisotropies present in the initial state of the collision. We can quantify this initial anisotropy in terms of eccentricity coefficients, which characterize the harmonic decomposition of the transverse energy (or entropy) density profile. In this work, the eccentricity $\epsilon_n$ of a single event is computed from the energy density in the transverse plane via
\begin{equation}
	|\varepsilon_n|^2 = \frac{ \langle r^n \mathrm{sin}(n\varphi) \rangle^2 + \langle  r^n \mathrm{cos}(n\varphi) \rangle ^2}{\langle r^n \rangle},
	\label{eq:eccentricity}
\end{equation}
where $\langle \cdot \rangle$ denotes the energy density weighted average. Naturally, the eccentricity for a selection (fixed IP-value in our case) is given by $\langle |\varepsilon_n|^2 \rangle$
It is well understood that through pressure gradients, these spatial anisotropies are converted into corresponding flow coefficients $v_n$ in momentum space, leading to a strong correlation between $\varepsilon_n$ and $v_n$. The quantitative mapping between initial-state eccentricities and final-state flow observables provides crucial constraints on the transport properties of the QGP, such as its shear and bulk viscosities.

\begin{figure}[t]
	\centering
	\includegraphics[width=0.5\textwidth]{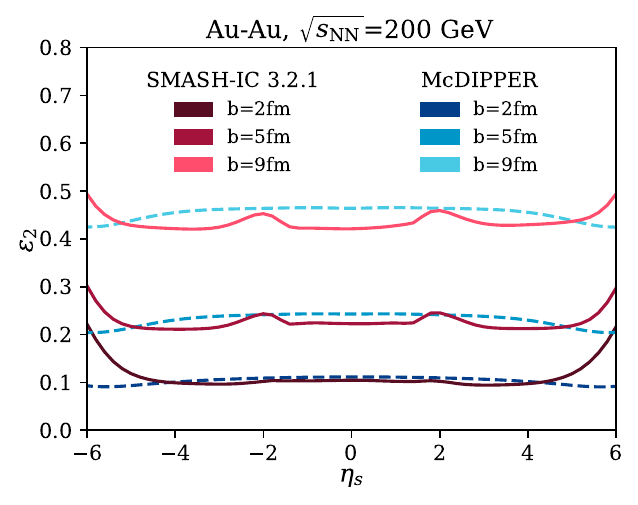}
	\caption[dE/d$\eta_s$]{Ellipticity as a function of spacetime rapidity $\eta_s$ in Au-Au collisions at $b=$~2~fm in {\Dipper} and the SMASH-IC at $\tau_{\mathrm{sw}}$ = 0.5~fm. To calculate $\varepsilon_2$, every event is recentered individually to its center of mass.}
\label{fig:ecc_eta}
\end{figure}

\cref{fig:ecc_eta} shows the ellipticity and triangularity as a function of spacetime rapidity from the two models. The most salient feature in the figure comes from the SMASH eccentricities, which increase at high rapidities. This is an artifact that arises from how the particle picture is translated to a smooth energy-momentum tensor. At high rapidities, SMASH finds itself depleted of particles, and the energy density decreases faster than the $n\neq 0$ harmonic integrals (the numerator in \cref{eq:eccentricity}). 
Additionally, slight increases in $\varepsilon_2$ can be observed at those points where the peaks in net-baryon and charge are located. This is a remnant of the fact that in SMASH, the hadronic degrees of freedom entangle energy and baryon(electric) charge deposition. In juxtaposition, the {\Dipper}, and likewise other partonic IC models such as Mc-EKRT \cite{Niemi:2015qia, Kuha:2024kmq}, will be dominated by gluon physics, dictating thus the behavior of quantities such as $\varepsilon_n$\footnote{It is important to state that the {\Dipper} does in fact present a relation between baryon stopping and energy density as it also includes the energy density arising from the deflection of quarks in the projectiles on the targets. There is a non-trivial portion $\sim 20-30\%$ of the energy density that is derived from this source and becomes relevant around the rapidity at which baryon stopping becomes maximal (see Figures 2 and 5 of Ref. \cite{Garcia-Montero:2023gex}). However, for the eccentricities, this contribution is generally drowned by the gluon energy deposition, and hence this latter source is the dominant behavior.}. For this reason, the $\varepsilon_n$ exhibit a concave behavior, with a relatively large quasi-flat plateau in the mid-rapidity region. If we think of reference flow as a hydrodynamical linear response to the initial conditions, $v_n = k_n \epsilon_n$, and hence the {\Dipper} comes closer to the general shape of the data~\cite{Garcia-Montero:2025bpn}. However, since outside of the midrapidity window $\eta_s$ and $\eta$ are not well matched to each other, this approximation does not function as well. Furthermore, the 3+1D hydrodynamics and the further re-sampling of particles at freezeout will re-shuffle the features presented here. We leave a full comparison within a complete hybrid framework for future work.
 
It is interesting to note that the {\Dipper} generally produces slightly higher eccentricities than SMASH. This is a result of the constrained IP-parameter dependence arising from the ICs of the $D_{adj}(x,\xT,\kT)$, as well as its QCD evolution, which results in a more granular event-by-event resolution. The SMASH results on the other hand depend on the smearing parameters used in the calculations. We found that a finer smearing will lead to a slightly higher eccentricity. In Appendix \ref{sec:smearing}, the reader can find a detailed assessment of the smearing/nucleon size dependence on both models.

In summary, despite pronounced differences in the underlying energy-density structure at a microscopic level, both models yield similar eccentricity coefficients. This confirms the general intuition that $\varepsilon_n$ is mostly governed by global geometrical features of the initial state rather than by the detailed local structure of the energy deposition.
\begin{figure}[t]
	\centering
	\includegraphics[width=\linewidth]{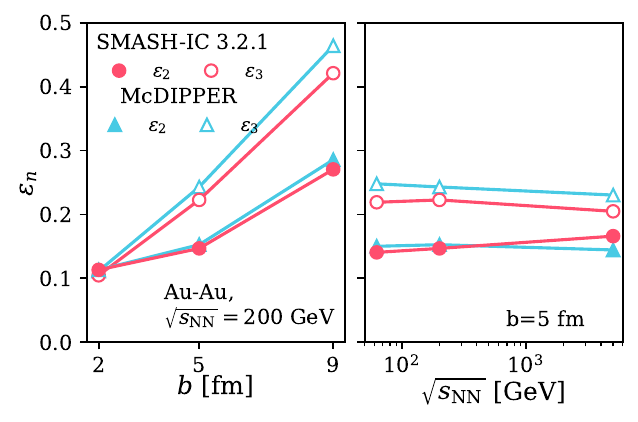}
	\caption[Eccentricity as function of b and sqrts]{Impact parameter (left) and energy dependence (right) of eccentricity harmonics $\varepsilon_2$ and $\varepsilon_3$ at midrapidity in the SMASH-IC and {\Dipper}. The left plot shows results from Au-Au collisions at $\sNN=$~200~GeV and the right plot shows the energy dependence of the $b=5$~fm collision system.} 
	\label{fig:ecc_impact}
\end{figure}
\section{Conclusion \& Outlook}
\label{sec:conclusions}
In this work, we compared to different initial state models for heavy-ion collisions, namely the string excitations based SMASH-IC and saturation based {\Dipper}, over a broad range of collision energies to assess their respective domains of applicability. We find that the two models show similar energy deposition $dE_\perp/d\eta_s$ at midrapidity for $\sqrt{s_{\mathrm{NN}}}=$~62.4 GeV suggesting that the two models have an overlapping region of applicability in this intermediate energy regime. However, with higher collision energies, {\Dipper} deposits substantially more transverse energy than SMASH. This reflects the importance of gluon- and small-$x$ dynamics, which are incorporated in the saturation framework but absent the string-based IC model.

Regarding the deposition of conserved charges, we find that the models differ not only in the position of the fragmentation peaks but also in the total amount of charge and baryon number that is scattered from the initial nuclei. In particular, at higher energies, the SMASH initial condition stops many more baryons in the midrapidity region than experimentally expected. This highlights that {\Dipper} can describe baryon stopping consistently over a large energy range, while SMASH eventually breaks down at higher energies. It is clear that the modeling of string excitations and their transition from soft to hard processes plays a crucial role in determining baryon stopping within hadronic transport approaches. Furthermore, we show that this is strongly affected by the treatment of leading hadrons within the handling of string excitations and might change significantly if one changes it from the current cross section scaling to a formation time scaling, where the leading hadrons are allowed to interact only after a shortened formation time.

We want to emphasize the need for experimental data in the forward/backward fragmentation region of the collision to better understand the dynamics responsible for the stopping of baryons in the initial collisions of nucleons. Electromagnetic probes such as dileptons and photons, which escape the medium without strong interactions, could provide crucial insight into these early-stage dynamics~\cite{Coquet:2023wjk,Coquet:2021lca,Berges:2017fsa,Garcia-Montero:2019kjk,Garcia-Montero:2024lbl,Scheid:2025gew,Wu:2024pba,Garcia-Montero:2019vju}. It remains to be seen how these differences impact the final state observables after the hydrodynamic and hadronic evolution.  
\section*{ACKNOWLEDGEMENTS}
The authors want to thank Renan Góes-Hirayama and Carl B. Rosenkvist for helpful discussions.
 This work is supported by the Deutsche Forschungsgemeinschaft (DFG, German Research Foundation) through the CRC-TR 211 ``Strong-interaction matter under extreme conditions'' Project No. 315477589--TRR 211. OGM is supported by the European Research Council under project ERC-2018-ADG-835105 YoctoLHC, and by Maria de Maeztu excellence unit grant CEX2023-001318-M. Computational resources have been provided by the GreenCube at GSI. 
\appendix
\section{Leading hadrons and processes in SMASH}
\label{sec:leadinghadrons}
At high energies, hard string excitations in SMASH are hadronized using {\Pythia}'s full MPI framework and all the resulting particles of the interaction are put into the evolution of SMASH at the same space-time point. However, in the Lund string model picture, the newly produced hadrons do not form into physical hadrons immediately after the collision, but rather as the string expands and breaks. Consequently, their hadronic cross section in the evolution of SMASH is set to zero for the duration of some formation time (1fm by default), where-after they can interact with other particles. An exception to this is so-called leading hadrons, i.e., hadrons that contain valence quarks of the initial colliding hadrons, located at the string ends in momentum space. Because some of their contained quarks exist already at the time of the collision, the cross section of these leading hadrons should not be  zero during the formation time. For this reason, leading hadrons are treated differently: instead of setting their cross sections to zero, SMASH assigns them a reduced but non-zero cross section during the formation time. This reduction is proportional to the number of valence quarks it contains. For example, the cross section of a leading baryon that contains a valence diquark gets scaled down by a factor of 2/3, while that of a leading meson containing a quark is scaled by 1/2. 

\begin{figure}[t]
	\centering
	\includegraphics[width=0.5\textwidth]{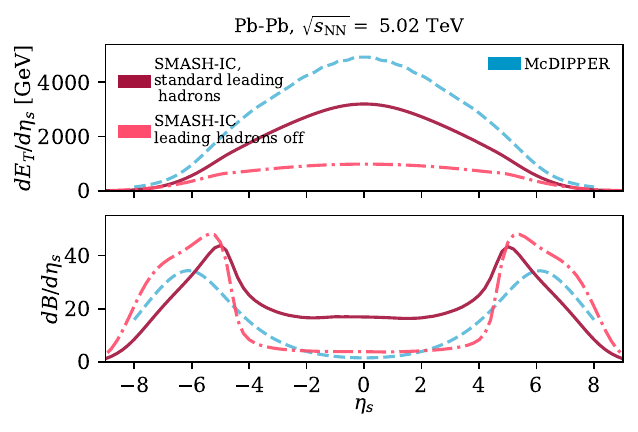}
	\caption[Impact of leading hadrons]{Longitudinal deposition of energy (top) and baryon number (bottom) from {\Dipper} and the SMASH-IC. The red and dashed line shows the effect of disabling leading hadrons inside SMASH, i.e., setting the cross section of all fragmented hadrons from string excitations to zero. Also shown are the results for disabling elastic and soft string interactions.}
\label{fig:leading_hadrons}
\end{figure}

This distinction has important consequences for baryon stopping. Because leading hadrons have a nonzero cross section, they can interact immediately after the initial nucleon–nucleon collision. This means they can undergo further interactions with other incoming nucleons, thus deflecting them to smaller (absolute) rapidities. 
In fact, at high energies, these secondary interactions between beam-nucleons and leading hadrons are responsible for most of the baryon stopping in SMASH. This is evident in \cref{fig:leading_hadrons}, where we compare the default leading hadron cross section scaling in SMASH to a setting where the cross section of leading hadrons is set to zero. One can see that disabling the leading hadrons significantly reduces the baryon deposition at midrapidity, while enhancing it at forward and backward rapidities. However, this also drastically reduces the transverse energy, indicating that in SMASH the transverse energy is strongly tied to the incoming baryons. This means that transferring energy into the transverse direction is coupled to stopping baryons, making it difficult to deflect energy without deflecting baryons away from their original beam rapidity.
This helps us to understand the constant baryon number density as a function of collision energy above 200~GeV that is observed in~\cref{fig:rapidity_shift}. Since leading hadrons can interact immediately with a non-zero cross section, they inevitably interact with the effectively opaque wall of incoming nucleons, regardless at which rapidity the nuclei pass each other. Therefore, the amount of baryons that are stopped in this way by leading hadrons does not depend on collision energy. In conclusion, the treatment of leading hadrons in the string interactions plays an important role in correctly describing the experimental data towards higher collision energies.

\section{Smearing dependence of observables}
\label{sec:smearing}
\begin{figure}[t]
	\centering
	\includegraphics[width=0.5\textwidth]{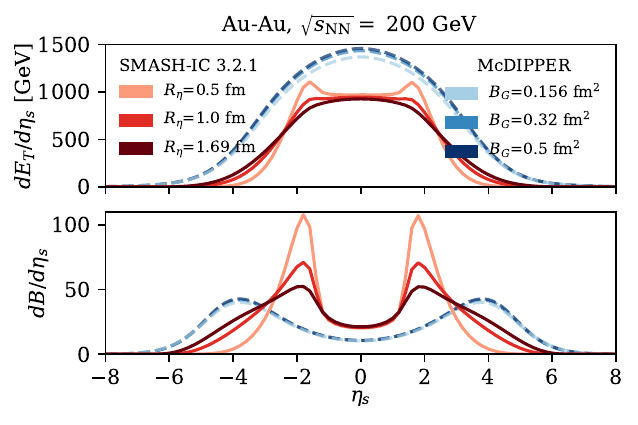}
	\caption[Impact of longitudinal smearing parameter]{Longitudinal deposition of energy (top) and baryon number (bottom) in Au-Au collisions at $\sNN=$~200~GeV and $b$=2~fm for different values of the longitudinal smearing parameter $R_{\eta}$ in the SMASH-IC and $B_G$ in {\Dipper}.}
	\label{fig:smearing_impact}
\end{figure}

In this Section, we want to explore how the results of this work depend on the respective smearing parameters of the models. \cref{fig:smearing_impact} shows the impact of the longitudinal smearing parameter $R_\eta$ on energy and baryon deposition. This figure is quite interesting, as it shows how the associated fluctuation sizes in both models are quite different, even for functionally similar "smearing" functions. In the case of the SMASH ICs, the midrapidity region is only slightly affected by it, while the size of the baryon deposition peaks increases significantly when $R_\eta$ becomes small. 
 On the other hand, the {\Dipper} only presents a transverse smearing parameter, $B_G$, as the longitudinal structure is given is emergent due to the dynamics of the collisions. Furthermore,  $B_G$ represents an effective smearing parameter for the incoming nucleons, not for the produced particles, marking another important difference. In the CGC, the dependence particle production is very non-linear due to the insertion of the nucleon density in \cref{eq:IPSatDip}\footnote{Our parameter $B_G$ is in fact highly constrained by DIS data at HERA~\cite{Albacete:2010sy}, we mean this appendix as a instructive piece.}. Hence, the total (now gluon) multiplicity is sensitive to the smearing size, or in other words, the effective nuclear density at momentum of collision. The reader can intuitively understand this in a more pictoric way. The inside of the exponential on the dipole of \cref{eq:IPSatDip} defines the so-called saturation scale of the system, so that 
\begin{equation}
	D(x,r) \sim e^{-r^2 Q_s^2 /4 }\quad \text{for small } r.
\end{equation} 
In momentum space, this distribution is highly peaked around $p_\perp\approx Q_s$. Due to kinematics, the gluons produced probe the left and right-coming targets at $x_{1,2}= p_\perp e^{\pm y}/\sNN$, which in conjunction with the simplification above gives a probed Bjorken-$x$ in the targets, 
 \begin{equation}
 	\langle x_{1,2}\rangle = \langle Q_s\rangle e^{\pm y}/\sNN\,.
 \end{equation}
 
 Naturally, increasing $B_G$ reduces the effective local $Q_s$, which means we are probing smaller-$x$ values. In other words, a more smeared nucleon has in principle the same effect as increasing the collision energy, and hence we get a larger energy density. 
This effect has been noted in many works, where BK and JIMWLK evolution of the targets lead to an effective transverse smearing of the nucleon \cite{Schlichting:2014ipa,Singh:2023rkg,Garcia-Montero:2025hys}. 
\begin{figure}[t]
	\centering
	\includegraphics[width=0.48\textwidth]{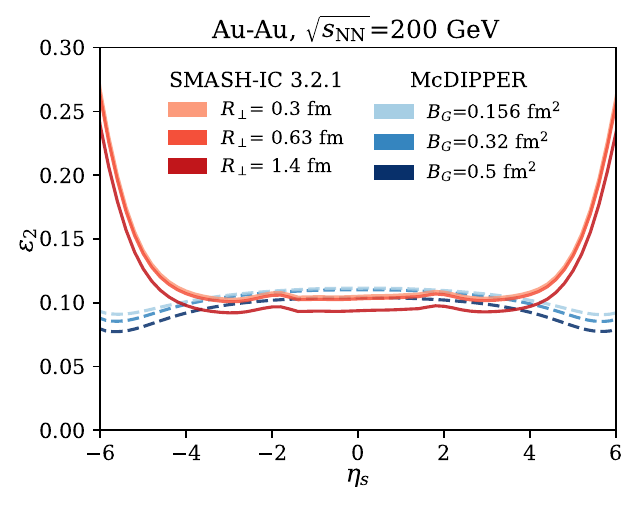}
	\caption[Impact of transverse smearing parameter]{Impact of the transverse smearing parameters $R_{\perp}$ and $B_G$ on the ellipticity $\varepsilon_2$ in Au-Au collisions at $\sNN=$~200~GeV and $b$=2~fm in the SMASH-IC and {\Dipper}.}
	\label{fig:ecc_smearing}
\end{figure}
\cref{fig:ecc_smearing} illustrates the impact of the transverse smearing parameter $R_\perp$ on the ellipticity. The two models show now similar trends. For small values of $R_\perp$, the SMASH-IC model gives a slightly larger $\varepsilon_2$. Similarly, {\Dipper} produces higher $\varepsilon_2$ values when $B_G$ becomes larger, i.e., the thickness function \eqref{eq:thickness} is more strongly smeared. $B_G$ in {\Dipper} and $R_\perp$ in SMASH are essentially different in that the former smears the size of individual nucleons, while the latter smears many particles that were created in the collision. Nevertheless, both have a similar effect on the created eccentricity: larger transverse size smears out fluctuations, while narrower smearing makes them more pronounced.  However, in SMASH ICs particles have some additional time to spread in random directions, which explains why $R_\perp$ has a relatively smaller effect on $\epsilon_2$, which seems to be determined mostly by the actual position of the created particles.  

 
\bibliography{References}
\end{document}